# Weak ferromagnetic insulator with huge coercivity in monoclinic double perovskite La$_2$CuIrO$_6$

Xingyu Zhang[1(†)], Bin Li[2(†)], Jie Cheng[2], Xu Chen[1], Lining Wang[1], Zilong Miu[1], Ziwan Song[1], Fengfeng Chi[2], Shengli Liu[2(*)], Z. H. Wang[3(*)]

[1]College of Electronic and Optical Engineering, Nanjing University of Posts and Telecommunications (NJUPT), Nanjing, Jiangsu 210023, China

[2]New Energy Technology Engineering Laboratory of Jiangsu Province & School of Science, Nanjing University of Posts and Telecommunications (NJUPT), Nanjing 210023, China

[3]School of Physics, National Laboratory for Solid State Microstructures, Nanjing University, Nanjing, Jiangsu 210093, People's Republic of China

(†) These authors contribute equally.

(*) Corresponding authors: liusl@njupt.edu.cn (S. Liu), zhwang@nju.edu.cn (Z. H. Wang)

Abstract

   Insulating ferromagnets with high T$_C$ are required for many new magnetic devices. More complexity arises when strongly correlated 3$d$ ions coexist with strongly spin-orbit coupled 5$d$ ones in a double perovskite. Here, we perform the structural, magnetic, and density functional theory (DFT) study of such double perovskite La$_2$CuIrO$_6$. A new $P2_1/n$ polymorph is found according to the comprehensive analysis of x-ray, Raman scattering and phonon spectrum. The magnetization reveals a weak ferromagnetic (FM) transition at $T_C$ = 62 K and short range FM order in higher temperature range. A huge coercivity is found as high as $H_C$~11.96 kOe at 10K, which, in combination with the negative trapped field, results in the magnetization reversal in the zero field cooling measurement. The first principle calculations confirm the observed FM state and suggest $La_2CuIrO_6$ of this polymorph is a Mott insulating ferromagnet assisted by the spin-orbit coupling.

## I. INTRODUCTION

Transition-metal oxides (TMOs) are usually correlated electron systems that offer many multifunctional properties, such as superconductivity, colossal magneto-resistance, and multiferroicity[1-3]. Searching for magnetic TMOs has lasted for over 100 years because of the wide range of applications in microwave devices, permanent magnets, and spintronic devices[4-10]. Moreover, TMOs become a renewed focus since a record of highest magnetic order temperature ($T_C$~1060K) has been found in the 5d TMO insulator $Sr_3OsO_6$ very recently[11].

5d TMOs have drawn considerable research interest in the condensed matter physics recent years, due to their comparable energy scales of strong spin-orbit coupling (SOC), electron correlation, crystal field and exchange interactions. In such systems, many novel quantum phases have been realized[12, 13], such as spin-orbit coupled Mott insulators[14], topological Mott insulators[15], superconductivity[16, 17], Weyl semimetals[16], axion insulators[18] and quantum spin liquid phases[19]. An interesting scenario appears in the perovskite iridates, where the 5d orbitals of iridium are split into $e_g$ and $t_{2g}$ states by the strong crystal field. The $t_{2g}$ state further forms a $J_{eff}=1/2$ doublet and a $J_{eff}=3/2$ quartet as a result of large SOC[14]. For systems with $Ir^{4+}$ ions, the partially occupied $J_{eff}=1/2$ state has properties very different from that of the spin-only $s = 1/2$ state. In fact, the detailed interplay of the aforesaid comparable energy scales is still not clear and under investigation.

Double perovskite TMOs provide additional degrees of freedom to choose different transition metal ions over the normal perovskite family, which have huge variety of properties such as high Curie temperature[20, 21], high magnetoresistance[20], metal-insulator transition[22] and half-metal[23]. More complexity arises when strongly correlated 3d ions coexist with strongly spin-orbit coupled 5d ones in a double perovskite. For instance, among the $La_2BIrO_6$ compounds, B = Fe, Co, Ni are reported to be noncollinear magnetism[7, 24, 25], B = Mn is ferromagnetism[26], and *B* = Zn, Mg show canted antiferromagnetism with unconventional Kitaev interaction[27-30]. Thus, these compounds promise new ways to develop desired magnetic functional materials for advanced technological applications.

The situation of double perovskite $La_2CuIrO_6$ is extremely complicated and in high debate. First of all, different crystal structures are reported by various groups. It is found that $La_2CuIrO_6$ forms a monoclinic $P2_1/n$ (No. 14) space group with $β$ near $90^o$ in early reports[31-33] and a recent preprinted paper[34], while a triclinic $P\bar{1}$ (No. 2) space group is also observed by x-ray diffraction (XRD) and neutron diffraction experiments[35]. Density functional theory calculations show that the energetics of these two structures are comparable, with the energy slightly lower in $P\bar{1}$ structure than in $P2_1/n$, and the energy difference only being ≈2meV/f.u.[36]. Then, the magnetic behavior of $La_2CuIrO_6$ is also controversial. Magnetic susceptibility measurements find an AFM transition at $T_N$ around ~70K and a weak FM behavior below ~50K[32, 34, 35]. Neutron diffraction suggests a possible spin configuration with collinear AFM spin arrangement in every *ac* plane and mutually orthogonal spin orientations in neighboring planes[35]. On the other hand, various spin structures have been proposed based on first principle calculations, such as a canted AFM[34] and a C-type AFM with the Cu and Ir spins anti-paralleled in a given *ac* plan, while paralleled in out-of-plane[36]. Therefore, the crystal structure and the magnetic properties remain to be fully understood.

## II. EXPERIMENTAL AND COMPUTATIONAL DETAILS

Polycrystalline samples of $La_2CuIrO_6$ are synthesized through the conventional solid state reaction method[34, 35]. The starting materials are $IrO_2$ (99.99%), CuO (99.99%) and $La_2O_3$ (99.99%). These mixtures are sufficiently grinded for 12 h and heated in air at 900,1100,1150 $^o$C for 60 h by a cooling of the furnace at a rate of 300 $^o$C/h with several intermediate grindings. The phase of all samples is checked by XRD on a Bruker diffractometer with Cu Kα radiation. The structural investigations have been done by Rietveld analysis by using the EXPGUI program. Electronic Raman scattering is performed at room temperature for the samples with a confocal microscope Raman spectrometer (Horiba HR-800). Magnetic properties are determined through a Quantum Design Physical Properties Measurement System (PPMS). Electronic structure calculations with high accuracy are performed by the full-potential linearized augmented plane wave (FP-LAPW) method implemented with the WIEN2K code[37]. The generalized gradient approximation (GGA)[38] is applied to the exchange-correlation potential calculation. The muffin tin radii are chosen to be 2.37 a.u. for La, 1.65 a.u. for O, 2.0 a.u. for both Cu and Ir atoms. The plane-wave cutoff is defined by $R \cdot K_{max} = 7.0$, where R is the minimum LAPW sphere radius and $K_{max}$ is the plane-wave vector cutoff.

## III. RESULTS AND DISCUSSION
### A. Crystal structure

Powder XRD data of polycrystalline $La_2CuIrO_6$ are presented in Fig. 1. All the lines in this pattern could be indexed to the monoclinic $P2_1/n$ structure[31-34]. This monoclinic double perovskite structure is derived from the perovskite structure by alternatingly placing Cu and Ir at the *B* site so that Cu and Ir ions form the fcc lattice respectively. The resulted lattice parameters are listed in Table I from the Rietveld refinements. The obtained lattice parameters are different from those in $P2_1/n$ structure of $La_2CuIrO_6$, although the cell volume is comparable with that of the literature[33]. For example, β~126° is larger than ~87° in early works[32, 33], indicating a considerably tilted crystal cell resulting in a much longer c-axis parameter. The attempts to fit the XRD data with earlier $P2_1/n$ and $P\bar{1}$ structures (not shown for simplicity) have failed since very large values of $\chi^2$ are obtained. Hence, a new polymorph is found in this $La_2CuIrO_6$ double perovskite due to different growth conditions. In fact, polymorphs have been found in various oxides. For example, there are five identified polymorphs of $Ga_2O_3$ depending on various growth conditions, known as corundum, monoclinic, defective spinel, and orthorhombic phases[39]. Moreover, cation vacancies can weaken the interaction between Cs and $PbI_6$ octahedra in $CsPbI_3$, stabilizing the Cubic Perovskite Polymorph[40]. In addition, polymorphs are also found in double perovskite $Sr_2FeMoO_6$ driven by defects[41].

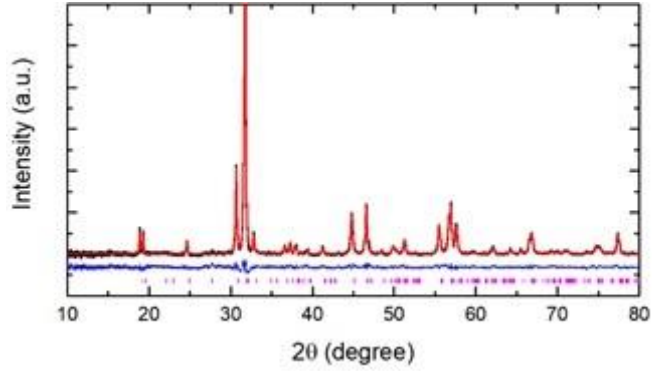

Fig. 1 (Color online): Observed, calculated, and difference profiles of Rietveld refined room-temperature XRD data of powder $La_2CuIrO_6$.

Table I. Structural parameters of obtained from Rietveld refinements of powder XRD data. Space group: $P2_1/n$, a = 5.60586(8) Å, b = 5.77985(9) Å, c = 9.57670(4) Å, β= 125.91 °, V=251.32(5) Å$^3$.

| Atom | x/a | y/b | z/c |
|---|---|---|---|
| La | 0.26489(9) | 0.45059(5) | 0.25385(4) |
| Cu | 0.500000 | 0.000000 | 0.500000 |
| Ir | 0.000000 | 0.000000 | 0.000000 |
| O1 | 0.17614(2) | 0.01098(4) | 0.24828(9) |
| O2 | 0.23012(3) | 0.69304(9) | 0.02923(4) |
| O3 | 0.34863(1) | 0.21278(1) | 0.07246(7) |
| wRp | 5.37% | $R_p$   3.84% | $\chi^2$   3.08 |

To verify this new polymorph, the phonon spectrum is calculated based on GGA according to the lattice constants from the above XRD Rietveld refinements. As shown in Fig. 2, there are no imaginary frequencies over the Brillouin zone and thus the $La_2CuIrO_6$ would be dynamically stable in this new polymorph structure. The calculated phonon dispersions indicate that there are 60 phonon bands extending up to ~550 cm$^{-1}$ and the point group is $C_{2h}$. $C_{2h}$ is Abelian group with 4 irreducible representations. The modes at Γ can be decomposed as $\Gamma = 18A_u \oplus 18B_u \oplus 12A_g \oplus 12B_g$, in which $A_u$ and $B_u$ modes are infrared active, $A_g$ and $B_g$ are Raman active. Thus, twenty-four Raman active modes are expected according to the calculation, as presented in Table II. To compare with experiments, electronic Raman scattering is carried out at room temperature. As shown in Fig. 3, four peaks at 205.8, 278.4, 385.9 and 530.3 cm$^{-1}$, labeled as $M_1$-$M_4$, dominate the spectrum. A comparison with the calculation suggests that peak $M_2$ is a mode with $A_g$ symmetry, peaks $M_3$ and $M_4$ are $B_g$ modes, and peak $M_1$ (hereby termed $M_1/M_1'$ ) is a superposition of an $A_g$ and a $B_g$ mode at nearby frequencies. Indeed, our calculations indicate $A_g$ modes at 209.3, and 272.3 cm$^{-1}$ and $B_g$ modes at 201.0, 386.2 and 522.4 cm$^{-1}$ in good agreement with the observed frequencies. The vibration representations of $A_g$ and $B_g$ modes are given in Fig. 4. Since these modes all correspond with Cu-O and Ir-O vibrations, only the patterns of the Cu/IrO$_6$ octahedra are given for clarity. The mode $M_2$ corresponds to an out-of-phase asymmetric stretching. While in mode $M_4$, vibration of CuO$_6$ octahedra corresponds to an out-of-phase symmetric breathing, and IrO$_6$ octahedra vibrates in an asymmetric stretching way.

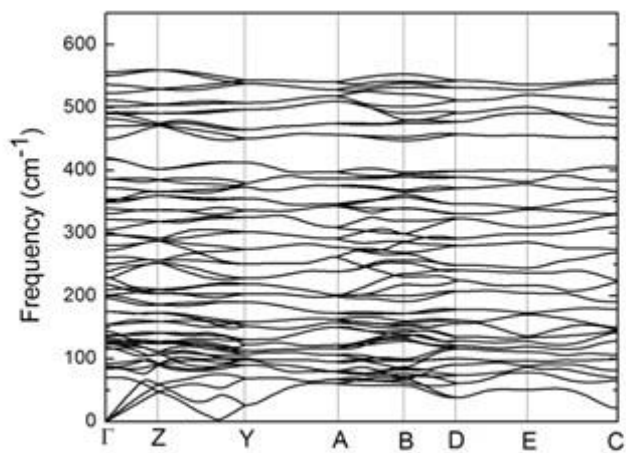

Fig. 2: Calculated phonon spectrum of La$_2$CuIrO$_6$.

Table II. DFT calculated frequencies of the Raman active modes represented in Fig. 2 and comparison with experimental data. The symbol "*" indicates the superposition of these two modes.

| Mode | Calc (cm$^{-1}$) | Observ. (cm$^{-1}$) | Mode | Calc (cm$^{-1}$) | Observ. (cm$^{-1}$) |
|---|---|---|---|---|---|
| $A_g$(1) | 92.6 | | $B_g$(1) | 87.6 | |
| $A_g(2)$ | 123.1 | | $B_g(2)$ | 127.0 | |
| $A_g(3)$ | 131.3 | | $B_g(3)$ | 143.7 | |
| $A_g(4)$ | 209.3 | 205.8* | $B_g(4)$ | 201.0 | 205.8* |
| $A_g(5)$ | 229.5 | | $B_g(5)$ | 227.9 | |
| $A_g(6)$ | 272.3 | 278.4 | $B_g(6)$ | 261.2 | |
| $A_g(7)$ | 305.1 | | $B_g(7)$ | 330.8 | |
| $A_g(8)$ | 338.9 | | $B_g(8)$ | 353.5 | |
| $A_g(9)$ | 350.5 | | $B_g(9)$ | 386.2 | 385.9 |
| $A_g(10)$ | 449.6 | | $B_g(10)$ | 482.0 | |
| $A_g(11)$ | 470.0 | | $B_g(11)$ | 490.5 | |
| $A_g(12)$ | 555.9 | | $B_g(12)$ | 522.4 | 530.3 |

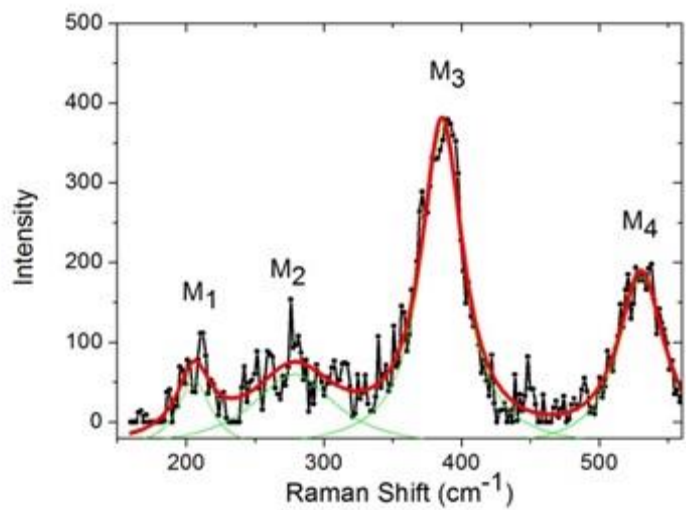

Fig. 3: (Color online) Raman spectra of $La_2CuIrO_6$ at room temperature. The solid lines are guides in the eyes.

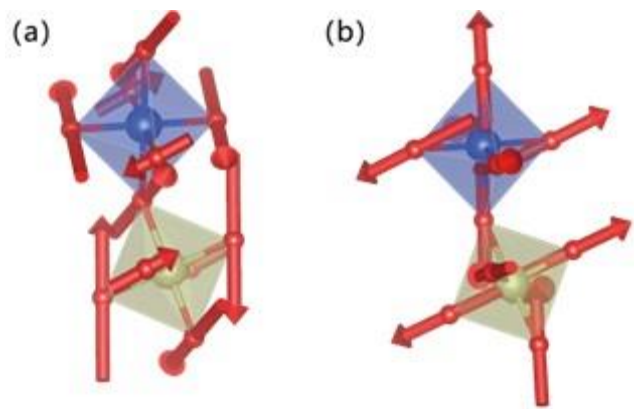

Fig. 4: (Color online) Vibration representations of the $Cu/IrO_6$ octahedra of modes: (a) $M_2$ mode ($A_g(6)$, 272.3 cm$^{-1}$); (b) $M_4$ mode ($B_g(12)$, 522.4 cm$^{-1}$). Blue circles: Cu; yellow circles: Ir; red circles: O.

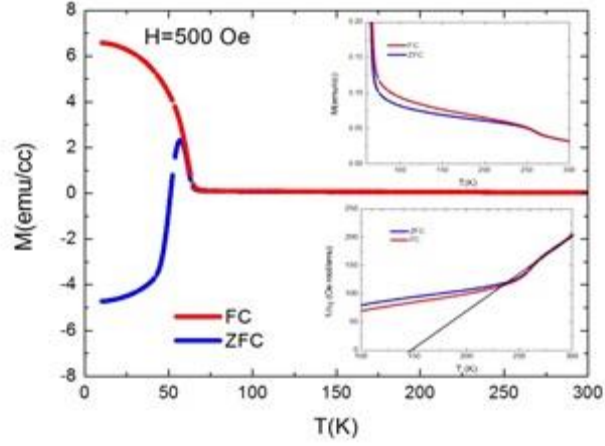

Fig. 5 (Color online) :Temperature dependent field-cooled (FC) and zero field-cooled (ZFC) magnetization of $La_2CuIrO_6$ at applied magnetic field H=500 Oe. Upper inset: enlarged plot of the main figure. Lower inset: Temperature dependent reciprocal susceptibility with straight line as the Curie-Weiss fit. Here $\Delta\chi = \chi - \chi_0$, where $\chi_0$ is a temperature independent contribution to $\chi$.

## B. Magnetic behavior

The magnetic behavior of $La_2CuIrO_6$ is very interesting. Double magnetic phase transitions in both $P\bar{1}$ and $P2_1/n$ structures are reported by literatures[34, 35], namely a paramagnetic to AFM phase transition at $T_N$ ~74 K and a weak FM transition below 60K. A similar kind of double transition behavior is also observed for $La_2ZnIrO_6$[29]. Fig. 5 illustrates the temperature dependent field cooled (FC) and zero field cooled (ZFC) magnetization (*M-T*) for our new polymorph of $La_2CuIrO_6$ sample. As the sample is cooled from the high temperature paramagnetic state, a sharp FM phase transition appears around $T_C$ = 62 K (determined from the peak of dM/dT data) with hysteresis between ZFC and FC curves. Another feature is that the ZFC magnetization shows a maximum at ∼57 K, goes through a zero point at the temperature $T_0$~50 K, and then remains negative down to the lowest temperature. The negative magnetization in the ZFC mode at low temperatures is noteworthy, which will be discussed later. Moreover, there is a distinct bifurcation between ZFC and FC curves when cooling below about 250 K as shown in the upper inset of Fig. 5 with an enlarged plot of the magnetization data, which can also be observed in the reciprocal susceptibility plot in the lower inset. Similar phenomena are observed in other ferro/ferri-magnets, such as $Pr_{1-x}Ca_xCoO_{3-\delta}$[42] and $Nd_{1-x}Ca_xCoO_{3-\delta}$[43], which is usually interpreted as a short-range FM order. The paramagnetic phase is further analyzed by plotting the temperature dependent inverse susceptibility in the lower inset of Fig. 5. The high temperature data, in the 260 K < $T$ <300 K window, nicely fit with the Curie-Weiss law yielding an effective magnetic moment of $\mu_{eff}$=2.54$\mu_B$/f.u. This is consistent with previous result of $P2_1/n$ structure [34] but larger than that of $P\bar{1}$ one [35]. Note that the obtained effective magnetic moment $\mu_{eff}$=2.54$\mu_B$/f.u. is very close to the ideal value $\sqrt{\mu_{Cu}^2 + \mu_{Ir}^2}$=2.449$\mu_B$ with $\mu_{Cu}$=$\mu_{Ir}$=$\sqrt{n(n+2)}\mu_B$ in the 'spin only' model if considering the spin configuration of $Cu^{2+}$/$Ir^{4+}$ ions. Interestingly the Curie-Weiss temperature is found to be $\theta_{CW}$ =+147K, consistent with the FM interaction.

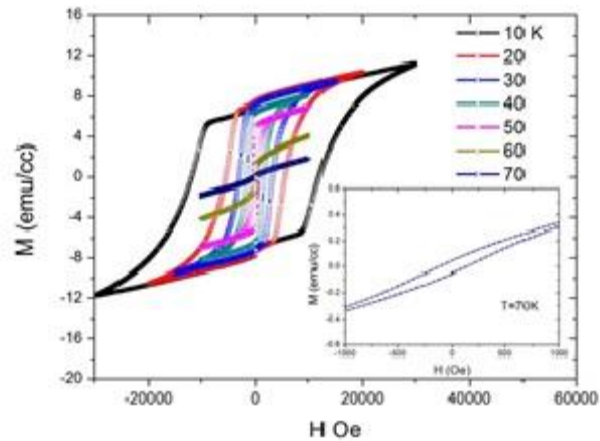

Fig. 6 (Color online): Magnetic hysteresis loops of $La_2CuIrO_6$ sample in a field range of -30 kOe to 30 kOe at temperature range from 10-70K. Inset: enlarged magnetic hysteresis loop at 70K.

To further understand the magnetic phases, magnetic isothermal (*M-H*) measurements are performed at several temperatures in Fig. 6. In earlier reports[34, 35], it is found that the magnetization of $La_2CuIrO_6$ system with both $P\bar{1}$ and $P2_1/n$ structures exhibits a small but visible hysteresis loop with the coercivity $H_C$ equaling to several hundreds of Oe, and linearly increases with the applied magnetic field as usually seen in other AFM materials. On the contrary, the *M-H* curves show large hysteresis loops of our new polymorph, indicating obvious FM order. Note that the magnetic hysteresis loop can still be observed at T=70K as shown in the inset of Fig. 6, where the temperature is above the FM transition temperature $T_C$ = 62 K, consistent with the framework of short-range FM order. Particularly, a huge coercivity of $H_C$~11.96 kOe at 10K is found, which is larger than ~2.5 kOe, the highest value for hard magnetic ferrites, and is similar to a very recent $YMn_{0.5}Cr_{0.5}O_3$ system[44]. As the temperature increases, the hysteresis loop shrinks and $H_C$ decreases significantly to ~580 Oe at T = 50 K, as shown in the inset of Fig. 7. The huge coercivity could be relevant to the magnetocrystalline anisotropy, since uniaxial magnetocrystalline anisotropy usually corresponds with a higher coercivity. The magnetocrystalline anisotropy can be understood through the ratios of remanent to saturation magnetization $M_r/M_s$. The ratios are 0.50 and 0.83 for polycrystalline samples with uniaxial and cubic magnetocrystalline anisotropy[44]. Since it is hard to explicitly define the saturation magnetization from the M-H curves due to the moderate slope at high fields, the value corresponding to the end point of the irreversible part is chosen as the saturation magnetization $M_s$. The saturation magnetization (~11emu/cc or 0.15$\mu_B$/f.u. at 10K) is significantly smaller than that of typical magnetic metals like $Nd_2Fe_{14}B$ (~1280 emu/cc), and typical ferrites like $CoFe_2O_4$ (~430 emu/cc), but is comparable with the very recent reported $Sr_3OsO_6$ (~49 emu/cc at 1.9K) [11]. The temperature dependence of $M_r/M_s$ is displayed in Fig. 7. At T=10K, a $M_r/M_s$ value close to 0.64 can be seen, suggesting that the sample tends to be uniaxial anisotropy at low temperatures. With temperature increasing to 50 K, the value increases gradually to 0.77, still less than 0.83 for the cubic anisotropy. Similar results are found in $YMn_{0.5}Cr_{0.5}O_3$ system[44].

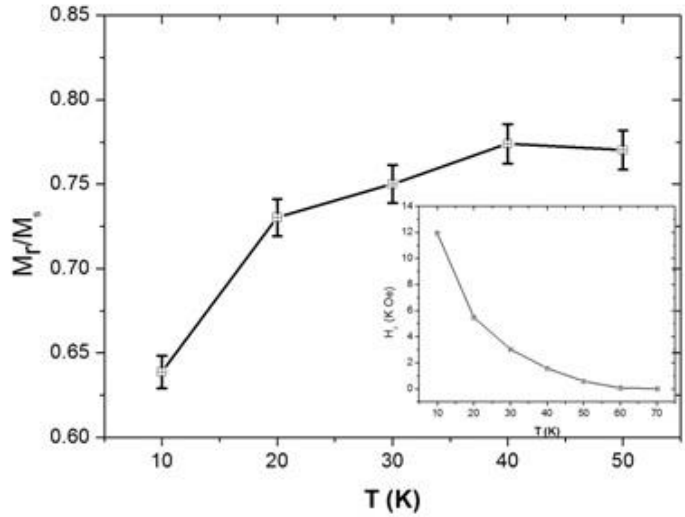

Fig. 7: The temperature dependence of $M_r/M_s$ of $La_2CuIrO_6$ sample. Inset: The temperature dependence of the coercivity.

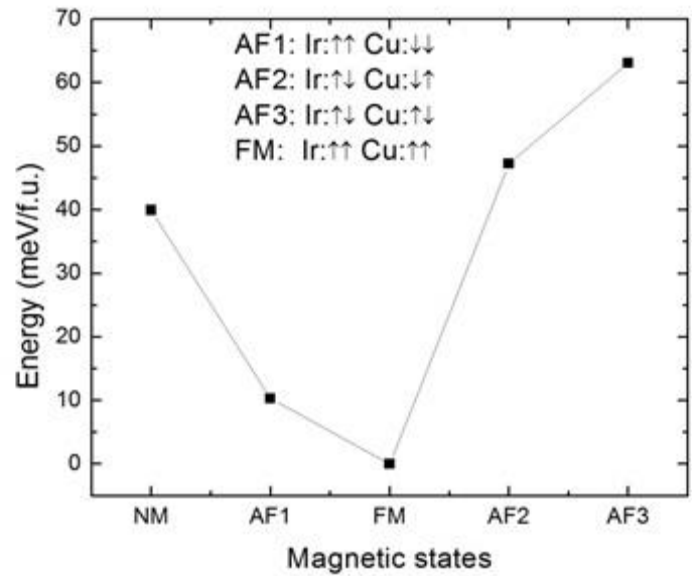

Fig. 8: Comparison of GGA total energy of $La_2CuIrO_6$ considering different magnetic structures. NM: nonmagnetism, AF: antiferromagnetism and FM: ferromagnetism.

The huge coercivity is also related to the magnetization reversal in the ZFC *M−T* measurement. The magnetization reversal, also called negative magnetization, is termed as a temperature dependent crossover of the dc magnetization from a positive value to a negative one (cooled under a positive applied magnetic field)[45, 46], which is different from a diamagnetic state that occurs in the case of superconducting or diamagnetic materials. The possible explanations for negative magnetization can be classified into different mechanisms, such as negative exchange coupling among ferromagnetic sublattices, among canted antiferromagnetic sublattices and among ferromagnetic/canted-antiferromagnetic and paramagnetic sublattices, etc[45]. The explanation based on the antiferromagnetic coupling between R-site and T-site cations can be excluded first in many $ABO_3$-perovskites with canted AFM sublattices residing at different crystallographic sites[47], since $La^{3+}$ ion is a nonmagnetic cation with fully occupied shells. For the second mechanism resulting from negative exchange coupling among ferromagnetic sublattices, the magnetization reversal always occurs in FC mode, such as $Co_2VO_4$[48]. However, the $La_2CuIrO_6$ sample shows magnetization reversal only in the ZFC mode, implying other causes in the present case. Actually, a similar negative magnetization only under ZFC mode in $YMn_{0.5}Cr_{0.5}O_3$ system has been presented very recently[44]. It has been found that the magnetization reversal in ZFC measurements is an artifact caused by negative trapped field of the superconducting magnet of the PPMS in combination with the huge coercivity. Considering that the PPMS superconducting magnet is usually turned off to zero from positive fields, the residual negative trapped field results in negative ZFC magnetization at low temperatures if the applied field is smaller than the coercivity. In fact, the coercivity is as high as 11.96 kOe at 10K for our $La_2CuIrO_6$ sample, much larger than the applied field H=500 Oe (see Fig. 5). With increasing temperature, the coercivity decreases rapidly to $H_C$~580 Oe at T=50 K (to see the inset of Fig. 7) comparable with the applied field, which results in the negative magnetization approaching to zero. With the further increase of the temperature the coercivity becomes smaller than the applied field, thus the ZFC magnetization becomes positive.

GGA calculations are performed to confirm the FM state in the new structure phase of $La_2CuIrO_6$ sample based on the lattice parameters listed in Table I deduced from the XRD measurements. Apart from the nonmagnetic (NM) state, FM structure with parallel alignment of all Cu and Ir spins is considered since $La_2CuIrO_6$ contains two magnetic ions. Three different AFM structures are considered, namely AF1, AF2, and AF3, possible within the unit cell of the new $P2_1/n$ structure phase. For AF1, the Ir-Ir spins and Cu-Cu spins are parallel coupling and Ir-Cu spins are antiparallel coupling. AF2 denotes in-plane antiparallel coupling of Ir-Cu spins but parallel coupling of

Ir-Cu spins between the layers. While AF3 denotes in-plane parallel coupling of Ir-Cu spins but antiparallel coupling of Ir-Cu spins between the layers. The energetics within GGA scheme of calculation is shown in Fig. 8. It can be seen that the FM structure is the lowest energetic spin configuration of our new structure phase. The AF1 structure lies 10.3meV higher than the FM state, resulting in a FM transition temperature of $T_C$=40K according to the mean-field estimation of $3/2\, k_B T_C = \Delta E/N$ with N=2 the number of magnetic ions in the formula unit, which is in reasonable agreement with the value ~62 K determined from M-T experiments. The calculated total magnetic moment is 1.15 $\mu_B$/f.u., with 0.69$\mu_B$ for Ir and 0.46$\mu_B$ for Cu respectively. Obviously, such value is a bit larger than that of the measured magnetization (0.15$\mu_B$/f.u. at 10 K). In general, DFT calculation would over-estimate the moment because the magnetic moment is defined by the integral inside the linearized augmented plane wave sphere, which is 2.0 a.u. for both Ir and Cu. Moreover, the DFT calculation is performed on a single crystal, while the magnetic measurements are carried out on polycrystalline samples. Such difference would also lead to the measured magnetization less than that of a full ferromagnetic alignment of the Cu and Ir cations.

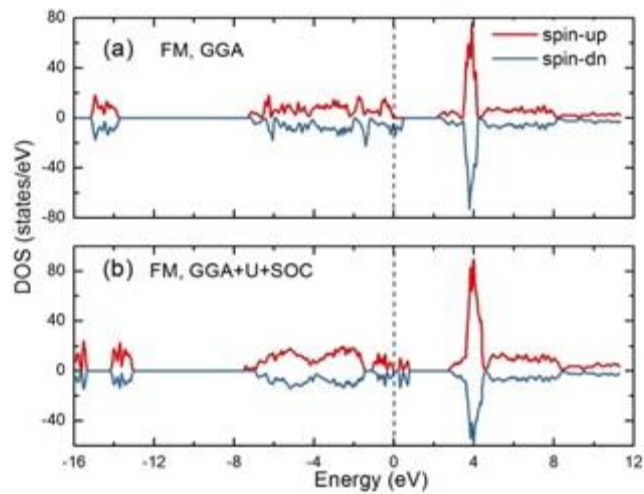

Fig. 9. (color online): The total density of states of the $La_2CuIrO_6$ system in two different situations: (a) FM GGA calculation, (b) FM GGA+U +SOC. The Coulomb interactions U=2 and 4 eV are adopted on Ir and Cu sites, respectively.

The total density of states (DOS) of the La$_2$CuIrO$_6$ system are presented in Fig. 9 for two different situations, namely the FM GGA and FM GGA+SOC+U calculation. The Coulomb interactions U=2 and 4 eV are adopted on Ir and Cu sites respectively. A sizeable gap of ~0.2 eV is observed only in the case of FM GGA+SOC+U, which indicates that our new *P*2$_1$/*n* structure phase of La$_2$CuIrO$_6$ is a Mott insulator assisted by the SOC effect. Note that similar result is reported in other *P*2$_1$/*n* and P2$_1$/*m* polymorphs with slightly higher gap of ~0.3 eV[34, 36].

In conclusion, we perform the structural, magnetic, and GGA study of the double perovskite La$_2$CuIrO$_6$. A new *P*2$_1$/*n* polymorph is found according to the comprehensive analysis of XRD, Raman scattering and phonon spectrum, which is different from the reported triclinic P$\bar{1}$ and monoclinic *P*2l/*n* as well. The magnetization reveals a weak FM transition at $T_C = 62$ K and short range FM order in higher temperature range. A Curie-Weiss fit of the inverse susceptibility yields $\theta_{CW}$=+147K and $\mu_{eff}$=2.54$\mu_B$/f.u., consistent with the 'spin only' model. The magnetic hysteresis loops indicate the magnetocrystalline anisotropy inclines to be uniaxial anisotropy. Particularly, a huge coercivity is found as high as $H_C$~11.96 kOe at 10K, which in combination with the negative trapped field results in the magnetization reversal in the ZFC measurement. The GGA calculations confirm the observed FM state and suggest that La$_2$CuIrO$_6$ of this polymorph is a weak ferromagnetic insulator assisted by the SOC.

## ACKNOWLEDGEMENTS

This work is partly supported by the National Natural Science Foundation of China (NSFC 11405089, 11504182, U1732126 and 11674054) and Natural Science Foundation of Jiangsu Province of China (No. BK20171440).